\documentclass[fleqn,10pt]{wlscirep}
\usepackage[utf8]{inputenc}
\usepackage[T1]{fontenc}
\usepackage{hyperref}
\usepackage{authblk}
\usepackage{xr}

\title{Machine Learning Detection of Majorana Zero Modes from Zero Bias Peak Measurements}
\author[1,2,*]{Mouyang Cheng}
\author[1,3]{Ryotaro Okabe}
\author[1,4]{Abhijatmedhi Chotrattanapituk}
\author[1,5,**]{Mingda Li}
\affil[1]{Quantum Measurement Group, MIT, Cambridge, MA 02139, USA}
\affil[2]{School of Physics, Peking University, Beijing 100084, China}
\affil[3]{Department of Chemistry, MIT, Cambridge, MA 02139, USA}
\affil[4]{Department of Electrical Engineering and Computer Science, MIT, Cambridge, MA 02139, USA}
\affil[5]{Department of Nuclear Science and Engineering, MIT, Cambridge, MA 02139, USA}

\affil[*]{e-mail: vipandyc@mit.edu}
\affil[**]{e-mail: mingda@mit.edu}

\begin{abstract}
Majorana zero modes (MZMs), emerging as exotic quasiparticles that carry non-Abelian statistics, hold great promise for achieving fault-tolerant topological quantum computation. A key signature of the presence of MZMs is the zero-bias peaks (ZBPs) from tunneling differential conductance. However, the identification of MZMs from ZBPs has faced tremendous challenges, due to the presence of topological trivial states that generate spurious ZBP signals. In this work, we introduce a machine-learning framework that can discern MZM from other signals using ZBP data. Quantum transport simulation from tight-binding models is used to generate the training data, while persistent cohomology analysis confirms the feasibility of classification via machine learning. In particular, even with added data noise, XGBoost classifier reaches $85\%$ accuracy for 1D tunneling conductance data and $94\%$ for 2D data incorporating Zeeman splitting. Tests on prior ZBP experiments show that some data are more likely to originate from MZM than others. Our model offers a quantitative approach to assess MZMs using ZBP data. Furthermore, our results shed light on the use of machine learning on exotic quantum systems with experimental-computational integration.

\end{abstract}
\begin{document}

\flushbottom
\maketitle

\thispagestyle{empty}

\section*{Introduction}
The identification of quantum many-body phases from experimental observations is one of the central tasks in condensed matter physics \cite{wang2023quantum,zhou2021high,wen2019experimental,von202040}. While symmetry-breaking phases can be detected unequivocally using local order parameters, topological phases of matter pose a more complex problem. Unlike the former, the topological phases cannot be characterized by local order parameters but instead carry global topological invariants \cite{wen2017colloquium}. As a result, detecting topological phases often requires an indirect measurement where topology can manifest, such as examining bulk excitations or specific boundary states \cite{qi2011topological}. Successful examples include the quantum anomalous Hall effect with insulating bulk and spin-polarized chiral edge states that can be probed by electrical transport \cite{yu2010quantized,chang2013experimental,deng2020quantum}, or topological Weyl semimetals with bulk Weyl fermions and surface Fermi arcs using photoemission \cite{armitage2018weyl}. In other cases, probing topology can become notably more challenging. In quantum spin liquids, for instance, bulk spinon excitations and edge Majorana fermions only leave subtle experimental evidence \cite{zhou2017quantum,wen2019experimental}. An enhanced capability to detect topological phases of matter will greatly enrich our understanding of quantum phases and hold paramount importance for next-generation microelectronic and quantum computing applications. 

Among the exotic topological phases of matter, Majorana Zero Modes (MZM), characterized by the non-Abelian, Ising-type anyonic statistics, have captured significant research and industrial attention over the past decade. Thanks to their unique ability to store information nonlocally, and their intrinsic zero energy that guards against hybridization, MZMs are deemed a highly promising platform to realize fault-tolerant topological quantum computation \cite{nayak2008non,kitaev2006anyons,alicea2011non}. Theoretically, MZMs were first proposed in the Kitaev 1D chain model with $p$-wave superconductor, where pairs of MZMs can emerge at the ends of the chain\cite{kitaev2001unpaired}. 
However, the evidence of $p$-wave superconductors has been elusive, with an unclear pathway to lift the double degeneracy of the spin pairing. Several remedies have been proposed. Fu and Kane suggest constructing MZMs using the proximity effect at the interface between an $s$-wave superconductor (SC) and a topological insulator, which resembles a $p_x+ip_y$ SC with additional time reversal symmetry \cite{fu2008superconducting}. 
Candidates like 5/2 fractional quantum Hall states\cite{read2000paired,moore1991nonabelions} and other platforms 
\cite{linder2010unconventional,ghosh2010non,alicea2010majorana,qi2010chiral,sato2009topological,sau2010generic}
are also potential candidates for hosting MZMs. Another milestone was reached to construct MZMs on a 1D nanowire with semiconductor (SM) coupled with proximity $s$-wave SC \cite{lutchyn2010majorana,oreg2010helical}. Under strong Rashba spin-orbit coupling and external Zeeman field, MZMs can emerge from an effective $p$-wave SC with the double degeneracy lifted. This SM/SC nanowire system has been considered extremely feasible to realize MZMs, with numerous experimental reports demonstrated in the past decade  \cite{das2012zero,test1,test2,test3,test4,test5,test6,test7,test8,test9,test10}. In these cases, the zero biased peaks (ZBPs) of the differential tunneling conductance under the scanning tunneling spectroscopy (STS) provide a strong experimental signature for MZMs \cite{jack2021detecting}. However, there has been a long concern that there are other topologically trivial states that can also produce ZBPs, such as Andreev bound states (ABS), Yu-Shiba-Rusinov states, or simply large disorders \cite{yin2015observation,pan2020generic,DasSarma_MZMPRR,frolov2020topological,valentini2021flux}. A systematic procedure to identify topological MZMs from experimental signals would be highly desirable.

In this work, we develop a machine-learning pipeline that aims to differentiate topological MZM from other topologically trivial states using experimental ZBP signals. The primary obstacles are the scarcity of experimental data and the absence of a universally acknowledged MZM ground truth. However, thanks to the STS technique, which can provide direct access to the single-particle density-of-states and further enables quantitative comparisons between experiments and computations, we were able to generate the ZBP training data computationally. Using effective Hamiltonian and quantum transport simulations, we cover a broad spectrum of physical parameters and mechanisms and further add data noises to mimic experiments. Although distinguishing MZM has created challenges due to the spectral similarity of ZBP between topological MZM and topologically trivial states, from a machine-learning perspective, this complexity is transformed into a classification task. Persistent cohomology analysis shows that the hidden global features of different topological classes remain robust, indicating that such a classification task is fundamentally machine-classifiable. By further implementing various machine-learning methods, such as linear classifiers, convolutional neural networks, and XGBoost, excellent accuracy is finally reached even with a reasonable level of data noise. We carry out additional tests on the experimental ZBP data from existing literature and found that some ZBP data are more likely to arise from MZM, while others are not. This does not rule out the potential presence of MZM in any of the reported experimental systems, given the limitation of the effective Hamiltonian approach and other experimental complexities not considered in this work. Our model offers an attempt to solve the MZM detection problem with machine learning. The work can also shed light on the application of machine learning in other exotic many-body quantum systems with very limited training data and a lack of ground truth.

\section*{Results}
\subsection*{Model setup}
\begin{figure}
  \centering
  \includegraphics[width=0.8\textwidth]{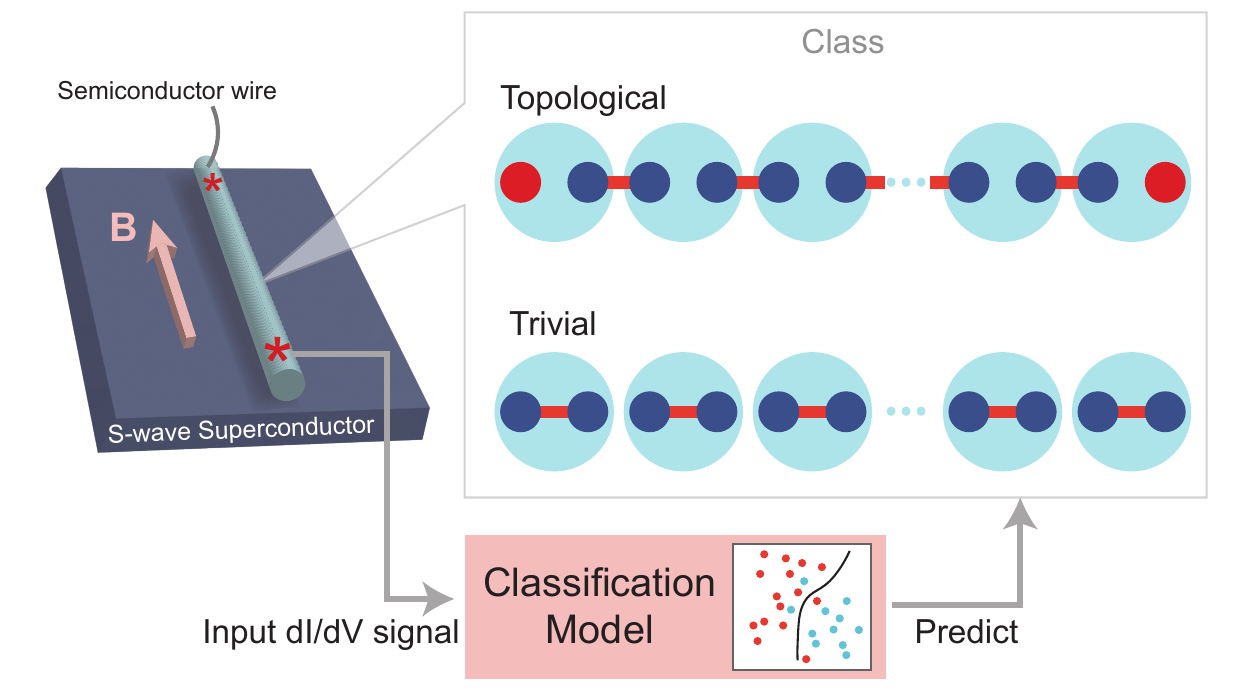}
  \caption{\textbf{The overview of the machine learning workflow to detect Majorana zero modes (MZMs) from zero bias peaks (ZBPs)}. The system consists of a 1D semiconducting nanowire coupled in proximity with an $s$-wave superconductor, which resembles a 1D $p$-wave superconducting Hamiltonian under a parallel magnetic field $B$. Training data are generated by an effective Hamiltonian approach. By modifying the on-site potential landscape, states that host topological MZMs and topologically trivial states are generated and labeled by the topological class. The tunneling conductance $dI/dV$ signals from the scanning tunneling spectroscopy are further computed using the tight-binding and quantum transport approach, which are used as input data. Various machine-learning models are established to achieve the MZM classification, with additional tests performed using existing experimental data.
  }
  \label{fig1}
\end{figure}

The general machine learning workflow is shown in Fig. \ref{fig1}. We consider the popular 1D SC/SM nanowire discussed earlier as the modeled system. The pristine nanowire system can be described by the 1D Boguliubov-de-Gennes (BdG) $p$-wave Hamiltonian $H=\frac{1}{2}\int \Psi^{\dagger}(x) H_{\rm{tot}}\Psi(x) dx$\cite{lutchyn2010majorana,sau2010generic,oreg2010helical}, where
\begin{align}
H_{\rm{tot}} 
&= T + H_{\rm{soc}} + U + H_{Z} + H_{\rm{couple}} \nonumber \\
&= \left(-\frac{\hslash^2}{2m^*} \frac{\partial^2}{\partial x^2} - i\alpha\frac{\partial}{\partial x}\sigma_y - \mu \right)\tau_z + V_Z\sigma_x + \Delta\tau_x.\label{Eq_Hamiltonian}
\end{align}
Here, $\hat{\Psi}(x)=\left(\hat{\psi}_{\uparrow}(x), \hat{\psi}_{\downarrow}(x), \hat{\psi}_{\downarrow}^{\dagger}(x),\hat{\psi}_{\uparrow}^{\dagger}(x)\right)^T$ spans a Nambu space with four spinors, and $\vec{\sigma}$ and $\vec{\tau}$ stand for Pauli matrices in the spin and particle-hole space, respectively. The five terms $T, H_{\rm{soc}}, U, H_{Z}$ and $H_{\rm{sc-sm}}$ denote the kinetic energy, spin-orbit coupling, on-site potential, Zeeman field coupling energy, and the SC-SM coupling, respectively. Detailed information about the choice of parameters is shown in Supplementary Information 1.

The Hamiltonian Eq. \ref{Eq_Hamiltonian} is the pristine Hamiltonian that leads to MZM. We further apply weak diagonal disorder $V_{\rm{imp}}(x) \sim \epsilon N(0,1)$ sampled from a normal distribution to mimic the noise but without destroying the topology.
In real experiments, trivial ZBPs may arise from a non-ideal potential landscape on the nanowire. At least two scenarios can lead to topologically trivial states, including \textbf{I.} quantum dots located at ends of the nanowire and \textbf{II.} large fluctuating disorder spread on the whole nanowire\cite{DasSarma_MZMPRR}. 
Therefore, for the topological trivial classes without MZM, we construct the Hamiltonian in two ways: 
For scenario \textbf{I.}, we add a Gaussian potential as an incommensurate on-site perturbation to the diagonal Hamiltonian. It has been shown that such smearing potential could be the culprit to the ABS. 
When the two Andreev bias peaks come closer under tuned parameters, these peaks will merge and form a trivial ZBP
\cite{stanescu2019robust,liu2017andreev}; For scenario \textbf{II.}, we amplify the disorder strength so that the fluctuation energy is comparable to the original chemical potential $\mu$. This can also give rise to topologically trivial states with ZBPs, creating a challenge for the MZM identification\cite{DasSarma_MZMPRR}. 

To generate the training data for machine learning, we cover a wide range of input Hamiltonian parameters (see Supplementary Information 1). The continuous Hamiltonian is discretized in real space to a finite tight-binding matrix. Then, we perform tight-binding simulations on this discretized system to calculate the tunneling conductance $G=\text{d}I/\text{d}V$ via the $S$ matrix formalism (see Methods for more details). A total of 12,000 labeled Hamiltonians are generated, with 4,000 for topological MZM, 4,000 for trivial ABS, and 4,000 for trivial large disordered states. The tunneling conductance signal can thus be calculated under sweeping a 2D parameter space composed of bias voltage $V_{\text{bias}}$ and Zeeman splitting $E_Z$, each with 28 different values. This leads to the use of $28\times 28$ image to represent the tunneling conductance data, labeled by either topological (hosting MZM) or trivial states (either ABS or large disorders) for machine learning classification. In addition, since some experimental works focus on 1D $dI/dV$ data without sweeping the Zeeman splitting, we single out the 1D data with zero Zeeman splitting for additional training. This can be done by searching the ZBPs while sweeping through $E_Z$ horizontally. Lastly, to improve the training robustness and bridge the theoretical-experimental gap,  we perform pre-processing on the raw data, including Gaussian smearing, additional noise, and anomaly detection on the dataset. More details on the Hamiltonian model, data generation, and processing can be found in Methods and Supplementary Information 1.

\subsection*{Global pattern with topological data analysis}
We first display typical tunneling conductance $dI/dV$ data generated from the workflow above in Fig. \ref{fig2}(b, c) for topological MZM and trivial classes (see Supplementary Information 2 for more examples). 
It can be seen that the 2D $dI/dV$ data from topological and trivial states have similar patterns. 
One earlier approach to achieve MZM pattern recognition \cite{DasSarma_MZMPRR} is finding the phase boundary between the topological and trivial classes. By pointing out the difference in the position of the topological phase transition compared to the pristine data, it was concluded that quantum dots and large disorder destroy the topology of the system, thus creating trivial ZBPs.
However, this approach is performed with fixed Hamiltonian parameters; when the parameters are unknown, discerning the topological MZM phase is still challenging for human eyes.

To investigate the potential intrinsic separability between the topological MZM and trivial classes, we employ the persistent cohomology analysis on a portion of the training dataset for all classes. Persistent cohomology is a type of topological data analysis (TDA) that studies the global feature difference at various scales. Figure \ref{fig2}(a) shows an example of persistent cohomology analysis on simplified 2D data. Starting from a gray image, a threshold value is tuned from the lowest pixel value to the highest. For a given threshold value, each pixel can be masked to binary black/white (lower/higher than threshold). Then two topological features emerge: Feature 0 identifies isolated black clusters in data (partially marked with light blue); Feature 1 focuses on closed loops encircled by a black cluster (partially filled with red). 
By sweeping the threshold values, different patterns assigned with different features emerge and annihilate, which create a birth-death scattered plot\cite{carlsson2020topological}. 
Therefore, persistent cohomology provides insights into the robustness and significance of these topological characteristics in the data.

Our analysis involves the 3D data composed of Bias voltage, Zeeman splitting, and other Hamiltonian parameters as one dimension. As a result, there is additional Feature 2 which captures voids or cavities entirely enclosed by surfaces. The persistent cohomology analysis is performed on our datasets using the GUDHI package with cubical complex \cite{gudhi:urm}. Results are shown in Fig. \ref{fig2}(d, e), where the difference between the topological MZM class and the trivial class can be seen clearly. Taking Feature 0 as an example; on the one hand, for the topological MZM dataset, there are very few clusters (light blue) that emerge near zero birth and annihilate early. On the other hand, for the trivial dataset, there is a continuous distribution of clusters that creates at zero birth and annihilates.

\begin{figure}[!htbp]
  \centering
  \includegraphics[width=\textwidth]{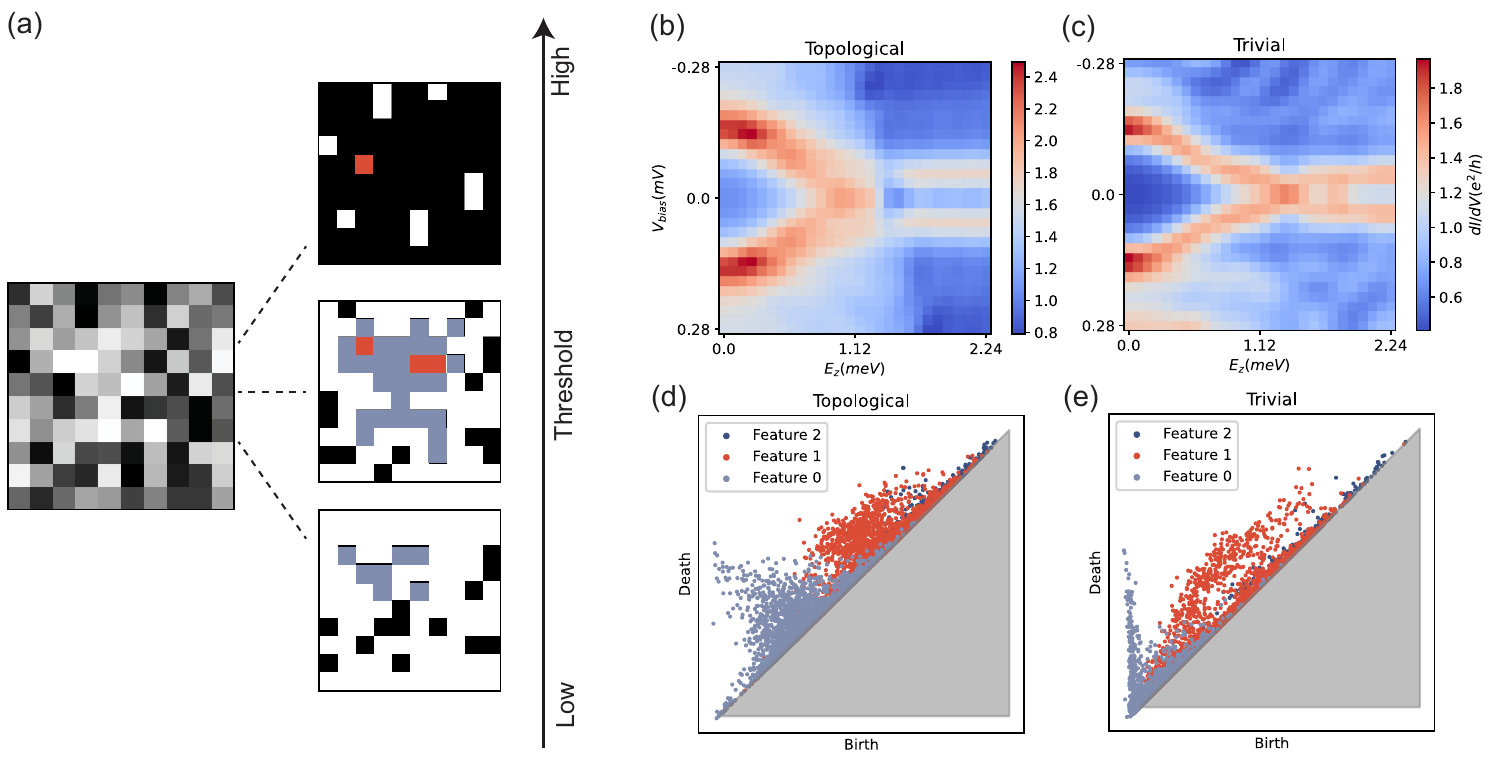}
  \caption{\textbf{Persistent cohomology analysis with the training data.} \textbf{(a)} Schematic of the principles on persistent cohomology, using simplified 2D data as an example. As the masking threshold is tuned from minimum to maximum magnitudes, features that mark isolated clusters (light blue) and surrounded loops (red) emerge and die out. Only features near a centered area are colored for better visualization.
  \textbf{(b, c)} Typical computationally generated tunneling conductance data used for machine learning training, for topological MZM and trivial classes, respectively. The 2D heatmap plots are tunneling conductance $dI/dV$ as a function of bias voltage $V_{bias}$ and Zeeman energy $E_z$; \textbf{(d, e)} Topological data analysis on topological MZM and trivial classes, respectively, using persistent cohomology analysis. Although the individual raw data in \textbf{(b, c)} are barely distinguishable with bare eyes, an obvious difference is shown between the topological MZM class and the other topological trivial classes through topological data analysis.
  }
  \label{fig2}
\end{figure}

Therefore, the persistent cohomology analysis implies that although the human eyes cannot readily distinguish the topological MZM states from trivial states, there exists a global topological feature difference between them. Such difference builds confidence that the MZM classification problem with ZBP is machine-separable prior to any design of machine-learning models.

\subsection*{Machine learning results}
\begin{figure}[!htbp]
  \centering
  \includegraphics[width=\textwidth]{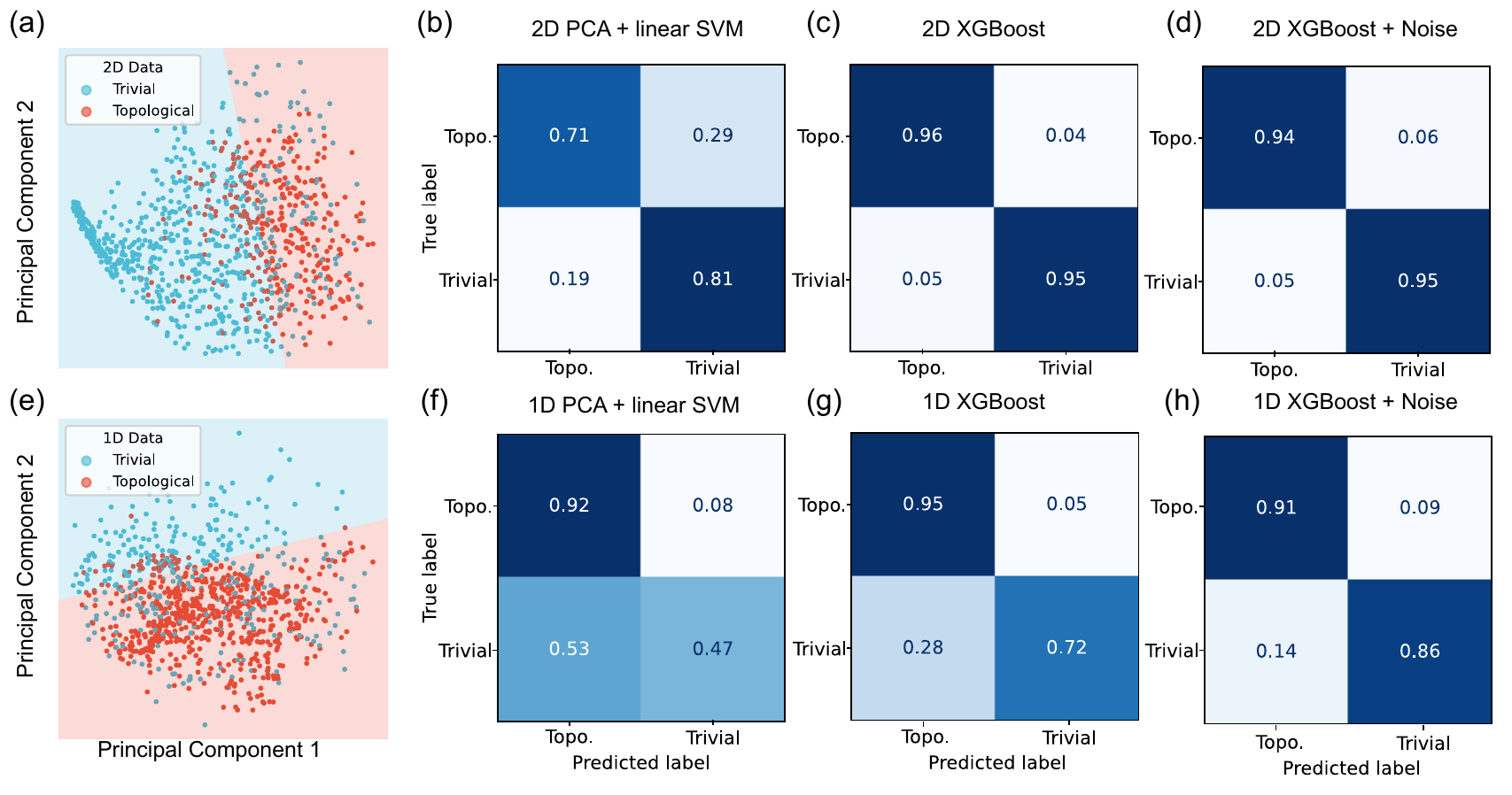}
  \caption{\textbf{Machine learning classifications to identify the topological MZMs using 1D and 2D tunneling conductance data.} \textbf{(a, e)} PCA analysis on the generated 2D (a) and 1D (e) data projected on the first two principal components. The SVM linear boundary roughly separates the topological MZM and trivial classes. \textbf{(b, f)} Confusion matrices for PCA + linear SVM learning results for 2D (b) and 1D (f) training data. \textbf{(c, g)} Confusion matrices from XGBoost for 2D (c) and 1D (g) training model without noise. \textbf{(d, h)} Confusion matrices from XGBoost for 2D (c) and 1D (g) training model with added data noise. Note that in all cases, the model with 2D data outperforms the model with 1D data, indicating the advantage of collect data with Zeeman energy sweeping.
  }
  \label{fig3}
\end{figure}

We employ a few machine-learning models to perform the topological MZM classification task. For the model inputs, 2D data of tunneling conductance images with $28\times 28$ pixels are flattened into 1D arrays, except for Convolutional Neural Network (CNN) which directly receives the 2D data.
As a baseline check, we first perform linear Principal Component Analysis (PCA) analysis to compress the data dimension. We reduce the 2D and 1D datasets' complexity to 2 dimensions for better visualization, and the reduced result with labels 0 or 1 are shown in Fig. \ref{fig3}(a, e). 
On the scattered plot for the first two leading principal components, there is no clear boundary between two separated clusters with different labels. Fig. \ref{fig3}(a, e) shows a linear Support Vector Machine (SVM) boundary line that separates two regions (shaded blue and red). However, there is a notable portion of data points crossing the boundary, indicating the limited power of linear classification at least on the PCA dimensionality-reduced dataset (performance shown in Fig. \ref{fig3}(b, f)). Particularly, for 1D PCA, the prediction of data labeled topological with 0.47 accuracy is close to random guess. Further attempts to use linear methods consistently provide lower accuracy than $90\%$ (see Supplementary Information 3), indicating the intrinsic data nonlinearity and calling on the necessity of nonlinear machine learning methods.

We carry out non-linear classification methods and ensemble methods including kernel-SVM, Random forest, CNN, and Extreme Gradient Boosting (XGBoost). The results as well as hyperparameters tuning process are described in Supplementary Information 3. Among them, XGBoost, which combines ensemble models and improved gradient boosting, gives overall better performance than other methods for both 1D and 2D tasks. The confusion matrix results for XGBoost training are shown in Fig. \ref{fig3}(c, d, g, h) as for 2D and 1D data, with and without data noise, respectively. It is worthwhile mentioning that binary classification with 2D tunneling conductance data for topological MZM class reaches $\sim 94\%$ accuracy, even in the presence of data noise. Additionally, although the 1D classifier gives a $\sim 28\%$ false positive for the topological MZM class, it still gives a high, $95\%$ confidence in true positive, and the overall accuracy still reaches a $85\%$. While adding noise reduces the accuracy of identifying trivial classes, it significantly improves the performance of detecting trivial classes from $\sim72\%$ to $\sim86\%$, which may be attributed to the large data variance and better data generalization.

The success in machine learning classification agrees well with the persistent cohomology observation. Also, the introduction of the Zeeman energy sweeping in 2D data outperforms the 1D data, indicating the benefit and possible necessity to take data with sweeping Zeeman energy. 

\subsection*{Experimental tests}
\begin{figure}[!htbp]
  \centering
  \includegraphics[width=0.8\textwidth]{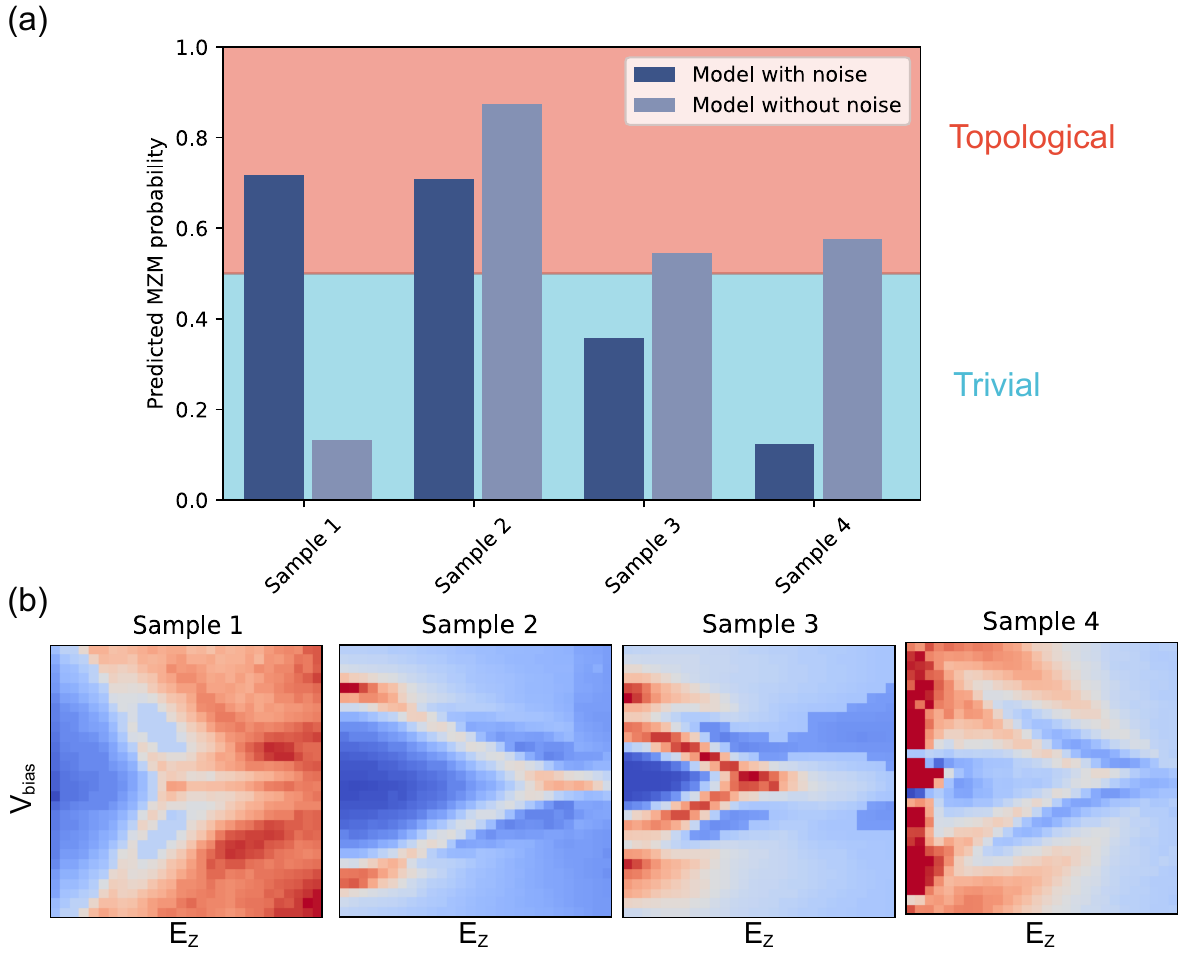}
  \caption{\textbf{Tests on experimental 2D ZBP data based on our machine learning models.} \textbf{(a)} The probability of positive prediction for the XGBoost model is plotted as a histogram. Both prediction results from the 2D XGBoost model with/without noise are shown for more information. Only 4 results labeled from sample 1 to sample 4\cite{test1,test5,test6,test8} imply that the system possibly hosts real MZM, while all the others suggest that they are unlikely to host MZMs. \textbf{(b)} Patterns of tunneling signal for sample 1 to sample 4 that pass the test.
  }
  \label{fig4}
\end{figure}

For the final part, we use our trained classifiers on real experimental ZBP data from recent literature. 
Since our classifier with 2D data input gives overall higher accuracy than the 1D classifier, we focus on the tests on the 2D ZBP data testing. Additional 1D data sets are shown in Supplementary Information 3. 
We extract 16 ZBP data images from 10 references during the past decade\cite{test1,test2,test3,test4,test5,test6,test7,test8,test9,test10}. The images are cropped online and processed to fit properly within our model input format (see Supplementary Information 4 for more details). Since XGBoost returns the continuous probability $p\in [0,1]$ before the final binary classification, here we show the probability since it carries more information than binary value, with a cutoff value $p_{\text{crit}}=0.5$. 
The positive result probability, i.e., the probability that the model suggests that the system hosts topologically MZMs, for the test set is shown in Fig. \ref{fig4}(a). Here we only emphasize the examples that manage to pass the trial test either with or without noise in the figure. 

Four experimental samples from four prior works pass the test from the 2D model either with or without data noise. The pattern of these samples are show in Fig. \ref{fig4}(b) from sample 1 to sample 4 in order\cite{test1,test5,test6,test8}. Among them, the most robust sample, upon which both models with and without noise imply positive MZM presence, has been retracted \cite{test5}. For the other 12 samples, the predicted probability for the existence of MZMs always lies consistently below 0.5, indicating that those systems are unlikely to host MZMs. The complete test results are shown in Supplementary Information 4. Overall, our model predicts that a dominant portion of experimental measurements is unlikely to host MZMs on SM-SC coupling nanowires. 

\section*{Discussion}
In this work, we propose a machine-learning pipeline to detect MZMs in experimentally measured differential tunneling conductance signals. Our work constructively aligns with the recent efforts to identify the topological MZMs from trivial states, replacing human eyes with machine-learning-based visual aids.  It offers a few potential advantages, including less bias and the possibility to quantify the performance. 

It is important to note that our model is only valid under a number of assumptions. It assumes that \textbf{I.} the experimental nanowire system can be well described within the 1D $p$-wave superconducting framework like the Kitaev chain; \textbf{II.} the physical mechanisms for impurity and disorder can be mimicked by modifying the diagonal potential landscape, and they are the only false positive sources for misleading ZBPs; \textbf{III.} the finite temperature effect can be modified by Gaussian smearing (see Supplementary Information 1).

To summarize, our work offers a framework as a machine learning attempt to identify MZMs from experimentally measured ZBP signals. Our classifier model could easily be generalized to suit other quantum property predictions, as long as the system can be well captured by effective model Hamiltonians. 
In the context of methodology, our machine learning model uses a mean-field approximation to capture the topological MZM feature under the condition of suppressed quantum fluctuations. This approach, to a broader aspect, could inspire more machine learning works integrated with experiments to tackle strongly correlated systems as a starting point. 
The model could also be further generalized to conduct parameters extraction on the experimental STS data, which resembles the philosophy of other machine-learning parameter extraction from experimental data such as time-resolved diffraction \cite{UED_Chen_2023} or neutron scattering \cite{Samarakoon2020}. 

\section*{Methods}
\subsection*{Tunneling conductance simulation}
The training data of our work is generated by tight-binding simulation on transport properties using the KWANT package\cite{groth2014kwant}.
To calculate the scattering matrix, we attach a normal SM nanowire with a lead to the end of the nanowire. The normal SM nanowire has the same form of Hamiltonian as the SC/SM system except for the SC coupling, i.e.
\begin{align*}
H_{\rm{normal}} 
&= T + H_{\rm{soc}} + U + H_{Z} \nonumber\\
&= \left(-\frac{\hslash^2}{2m^*} \frac{\partial^2}{\partial x^2} - i\alpha\frac{\partial}{\partial x}\sigma_y - \mu_{\text{normal}} \right)\tau_z + V_Z\sigma_x.
\end{align*}
Note that there is a finite difference between the normal wire and the SC/SM nanowire in a chemical potential $\mu_{\text{normal}}-\mu =eV_{\text{gate}}$, which represents the gate voltage added to the scattering region.
As for the lead, the on-site Hamiltonian is the same as the normal nanowire except for an additional potential barrier $V_{\text{barrier}}$:
\begin{equation*}
H_{\rm{normal}} = \left(-\frac{\hslash^2}{2m^*} \frac{\partial^2}{\partial x^2} - i\alpha\frac{\partial}{\partial x}\sigma_y - \mu_{\text{normal}} + V_{\rm{barrier}} \right)\tau_z + V_Z\sigma_x.
\end{equation*}
All relevant physical parameters in the Hamiltonian can be found listed in Supplementary Information 1. After constructing such a system, KWANT allows convenient calculation on the scattering matrix $S$ on the defined scattering region, i.e., the connecting junction on the lead. 
Then the tunneling conductance $G_0(E)=dI/dV$ with respect to energy $E$ in units of $e^2/h$ can thus be computed by\cite{liu2017role}:
\begin{equation*}
G_0(E)=2+\sum_{\sigma, \sigma^{\prime}=\uparrow, \downarrow}\left(\left|r_{e h}^{\sigma \sigma^{\prime}}\right|^2-\left|r_{e e}^{\sigma \sigma^{\prime}}\right|^2\right),
\end{equation*}
where $r_{e h}$ and $r_{e e}$ are the Andreev and normal reflection amplitudes from the $S$ matrix, respectively. The calculated tunneling conductance is energy-dependent, and by sweeping the Zeeman energy $E_Z$, we can obtain a diagram with $dI/dV$ versus $E_Z$ and bias energy(voltage) $V_{\rm{bias}}$, which finally gives an image of our 2D data.
Such numerical method is extensively performed in the relevant area of literature, and we refer readers to references like \cite{prada2012transport,liu2017role,pan2020generic,DasSarma_MZMPRR} for more details.

\subsection*{Data processing before training}
After generating these raw data, we add Gaussian smearing by adding Gaussian kernel transform to our 2D image:
\begin{equation*}
F(G) \sim \text{exp}(-G^2/2\sigma^2)
\end{equation*}
where $\sigma=1 \rm{pixel}^{-1}$. The reason for such processing are of two folds: First, such Gaussian smearing mimics the finite temperature effect of experimental measurements based on our zero-temperature simulation (see Supplementary Information 1);
Secondly, our smearing also smooths out the experimental STS measurement signal, mimicking the resolution function resembling the Gaussian kernel.

In addition to such smearing, to ensure the robustness of our model and emulate the measurement noise we further add a small noise to the tunneling conductance signal subject to the normal distribution $\delta G \sim 0.2N(0,1) e^2/h$. 

\subsection*{Methods of machine learning}
For the machine learning part, the methods and models we used are well implemented in Python open-source packages. We use the scikit-learn package\cite{scikit-learn} for PCA analysis, SVM, random forest, and XGBoost classification, and we implement the Pytorch\cite{paszke2019pytorch} package for building up the simple CNN network for classification. All methods are well-documented and conventional, and the model and hyperparameter settings are listed in Supplementary Information 3.

\section*{Data and code availability}
The data used in this study are numerically generated using our code implementing KWANT, and the code used in this study is available at \url{https://github.com/vipandyc/ML_majorana}. 

\section*{Acknowledgements}
The authors thank J Sau, B November, and J Hoffman for the helpful discussions. RO and AC acknowledge support from the US Department of Energy (DOE), Office of Science (SC), Basic Energy Sciences (BES), Award No. DE-SC0021940, and National Science Foundation (NSF) Designing Materials to Revolutionize and Engineer our Future (DMREF) Program with Award No. DMR-2118448. AC is partially supported by the MIT-QEC Seed Fund. ML acknowledges the support from the Class of 1947 Career Development Chair and support from R Wachnik.

\bibliography{bibliography.bib} 
\end{document}


\maketitle
\flushbottom
\tableofcontents
\thispagestyle{empty}
\section{Generation of dataset}
\subsection{Topological and trivial Hamiltonian}
For simulating the $dI/dV$ STM measurement from experiments in such SC-SM coupling wire, we start from the 1D p-wave pairing superconducting Hamiltonian coupled to a nanowire $H=\frac{1}{2}\int \Psi^{\dagger}(x) H_{\rm{tot}}\Psi(x) dx$, where
\begin{align}
H_{\rm{tot}} 
&= T + H_{\rm{soc}} + U + H_{Z} + H_{\rm{couple}} \nonumber\\
&= \left(-\frac{\hslash^2}{2m^*} \frac{\partial^2}{\partial x^2} - i\alpha\frac{\partial}{\partial x}\sigma_y - \mu \right)\tau_z + V_Z\sigma_x + \Delta\tau_x.
\end{align}
Here, $\hat{\Psi}(x)=\left(\hat{\psi}_{\uparrow}(x), \hat{\psi}_{\downarrow}(x), \hat{\psi}_{\downarrow}^{\dagger}(x),-\hat{\psi}_{\uparrow}^{\dagger}(x)\right)^T$ spans a Nambu space with four bands, and $\vec{\sigma}$ and $\vec{\tau}$ stand for Pauli matrices in the spin and particle-hole space, respectively. The five terms $T, H_{\rm{soc}}, U, H_{Z}$ and $H_{\rm{sc-sm}}$ describe the kinetic energy, spin-orbit coupling, on-site potential, Zeeman field coupling energy, and the SC-SM coupling.
In our model, the only difference between topological and trivial data lies in the on-site potential $U$. For the topologically MZM candidates, we modify the Hamiltonian as
\begin{equation}
H_{\rm{topo.}} 
= \left(-\frac{\hslash^2}{2m^*} \frac{\partial^2}{\partial x^2} - i\alpha\frac{\partial}{\partial x}\sigma_y - \mu + \epsilon V_{\rm{imp}}(x)\right)\tau_z + V_Z\sigma_x + \Delta\tau_x.
\end{equation}
Where $V_{\rm{imp}}(x) \sim N(0,1)$ follows the standard normal distribution, and $\epsilon=0.2 \rm{meV}$, which is typically much smaller than the chemical potential $\mu \sim 1 \rm{meV}$ in our simulation. We choose to generate the topological data with such noise added to our model to demonstrate that the real Majorana systems should be topologically robust to moderate exterior perturbation.

On the other hand, the topologically trivial MZM Hamiltonian is considered in two situations: \textbf{I.} Andreev bound states induced by inhomogeneous potential, which mimics quantum dots or impurities fixed at the normal lead:
\begin{equation}
H_{\rm{andreev}} 
= \left(-\frac{\hslash^2}{2m^*} \frac{\partial^2}{\partial x^2} - i\alpha\frac{\partial}{\partial x}\sigma_y - \mu + V_{dot}e^{-x^2/2\sigma^2} \right)\tau_z + V_Z\sigma_x + \Delta\tau_x.
\end{equation}
\textbf{II.} Large disordered states with disorder strength $W=1.0\rm{meV}$ comparable with the on-site chemical potential:
\begin{equation}
H_{\rm{disordered}} 
= \left(-\frac{\hslash^2}{2m^*} \frac{\partial^2}{\partial x^2} - i\alpha\frac{\partial}{\partial x}\sigma_y - \mu + W V_{\rm{imp}}(x) \right)\tau_z + V_Z\sigma_x + \Delta\tau_x.
\end{equation}

Fig. \ref{figs1} below shows the sketch of our nanowire alongside with different types of potential with different classes. 

\begin{figure}[!htbp]
  \centering
  \includegraphics[width=0.75\textwidth]{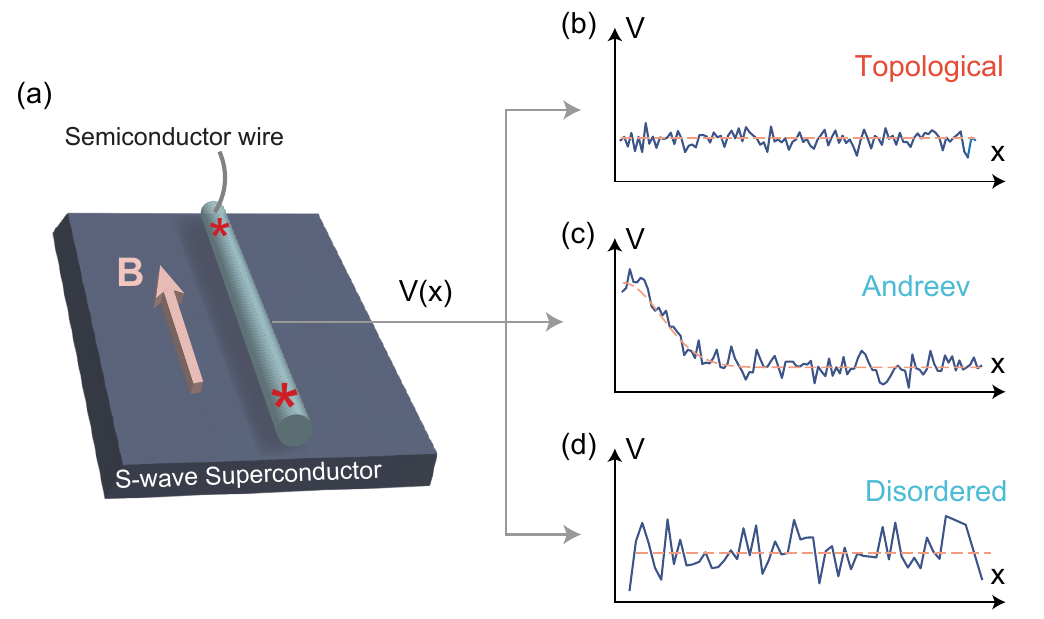}
  \caption{Schematic sketch of topological and trivial Hamiltonian. \textbf{(a)} The system consists of a 1D semiconducting nanowire coupled in proximity with an $s$-wave superconductor, which resembles a 1D $p$-wave superconducting Hamiltonian under a parallel magnetic field $B$. \textbf{(b)} Topological MZMs with low disorder, \textbf{(c)} trivial Andreev bound states and \textbf{(d)} trivial states with huge fluctuating disorder are simulated by modifying the potential landscape $V(x)$.
  }
  \label{figs1}
\end{figure}

\subsection{Choice of parameter settings}
To make our model more generalized, we sample constant parameters that characterize the nanowire system within a given range, which spans $50\%$ deviation from a reasonable value in literature \cite{DasSarma_MZMPRR}. The detailed parameter settings are listed in table \ref{table1_params} below.

\begin{table}[!htbp]
\centering
\begin{tabular}{lll} 
Parameter & Value range \\
\hline \hline 
Nanowire length $L$ [$\mu$m] & 1 $\sim$ 3 \\
SC chemical potential $\mu$ [meV] & 1 $\pm$ 0.5 \\
Lead scatter barrier [meV] & 10 $\pm$ 5 \\
SOC coupling $\alpha$ [eV/\si{\angstrom}] & $0.5 \pm 0.25$ \\
SC coupling $\Delta$ [meV] & $0.2 \pm 0.1$ \\
Max incommensurate potential $V_{dot}$ [meV] & $1.5 \pm 0.75$ \\
Effective mass $m^*$ & 0.015$m_e$ \\
SM chemical potential $\mu_0$ [meV] & 25 \\
\hline
\end{tabular}
\caption{Constants used in our simulation. For each data simulation, a nanowire is generated from random sampling within the range given in the table.}
\label{table1_params}
\end{table}

Note that to employ scattering calculation, as common practice, we attach a normal lead with semiconductor chemical potential $\mu_0$ to the end of the nanowire and calculate the scattering matrix $S$ using the KWANT package. The differential conductance $g=dI/dV$ can then be calculated via the formalism in the "Methods" section in the main text.

\subsection{Smearing and noise}

\begin{figure}[!htbp]
  \centering
  \includegraphics[width=1.0\textwidth]{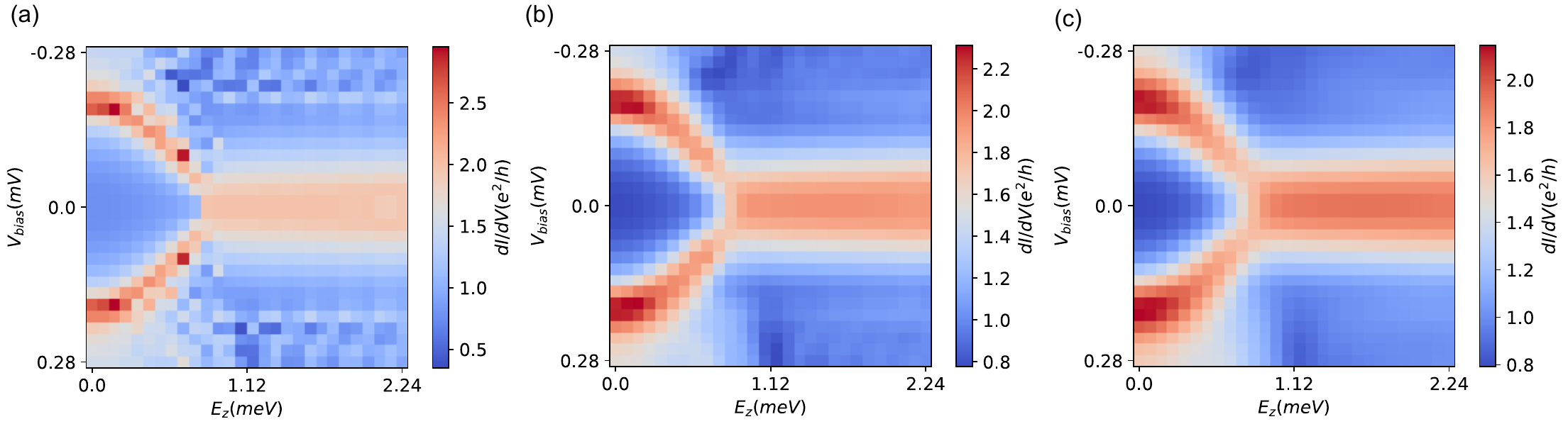}
  \caption{Example of generated data with different smearing. (a) Original data without smearing; (b) Gaussian smearing with $\sigma=1$ on the $28\times 28$ pixel; (c) Gaussian smearing with $\sigma=1.5$ on the $28\times 28$ pixel.
  }
  \label{figs2}
\end{figure}

While the Hamiltonian above simulates transport properties at zero temperature, the signals in experiments are in fact measured at low finite temperature, which is not completely consistent with our model. In principle the differential conductance at finite temperature $g_T(E)$ can be calculated via a convolution of zero temperature conductance $g_0(E)$:
\begin{equation}
g_T(E)=-\int_{-\infty}^{+\infty} g_0(E')\frac{\text{d}f(E-E')}{\text{d}E'} \text{d}E'
\end{equation}
Where $f(E-E')=1/(e^{\beta (E-E')}+1)$ is the Fermi distribution.
At zero temperature its derivative gives $\delta$ function, while for low finite temperature the modified differential conductance can be mimicked by applying a Gaussian smearing along the energy axis.
Fig. \ref{figs2} shows the smeared results of our original generated data. This smeared pattern matches very well with the results from previous literature\cite{liu2017role}.

Apart from mimicking the finite-temperature effect, we would like to stress that a 2D smearing also smooths out the experimental STM measurement signal, due to the resolution function analogous to the Gaussian kernel for observable $G$\cite{tersoff1985phys}:
\begin{equation}
F(G) \sim \text{exp}(-G^2/2\sigma^2)
\end{equation}
In addition to such smearing, to ensure the robustness of our model and emulate the measurement noise we further add a small noise to the $dI/dV$ signal subject to the normal distribution $\delta G \sim 0.2N(0,1) e^2/h$. To sum up, on one hand, adding smearing and additional noise is important in bridging the gap between computational results and real experimental data and making our model more realistic; on the other hand, such post-processing also enhances the robustness of our machine learning model, and makes it produce more reliable results.

\subsection{Extracing 1D training data from 2D images}
Now that we're ready with our 2D image dataset, we can extract 1D $dI/dV$ curves from these images.
Basically, 1D curves are legitimate $dI/dV$ versus bias voltage data with centered zero bias peak.
As an arbitrary choice to identify such peaks, all 1D curves with a centered value larger than its five vicinity pixels and larger than 1.8 $e^2/h$ are extracted and included in the 1D dataset.
This creates a much larger number of individual data than the previous 2D image set because there can almost always be multiple candidate 1D curves for a single image.

\begin{figure}[!htbp]
  \centering
  \includegraphics[width=1.0\textwidth]{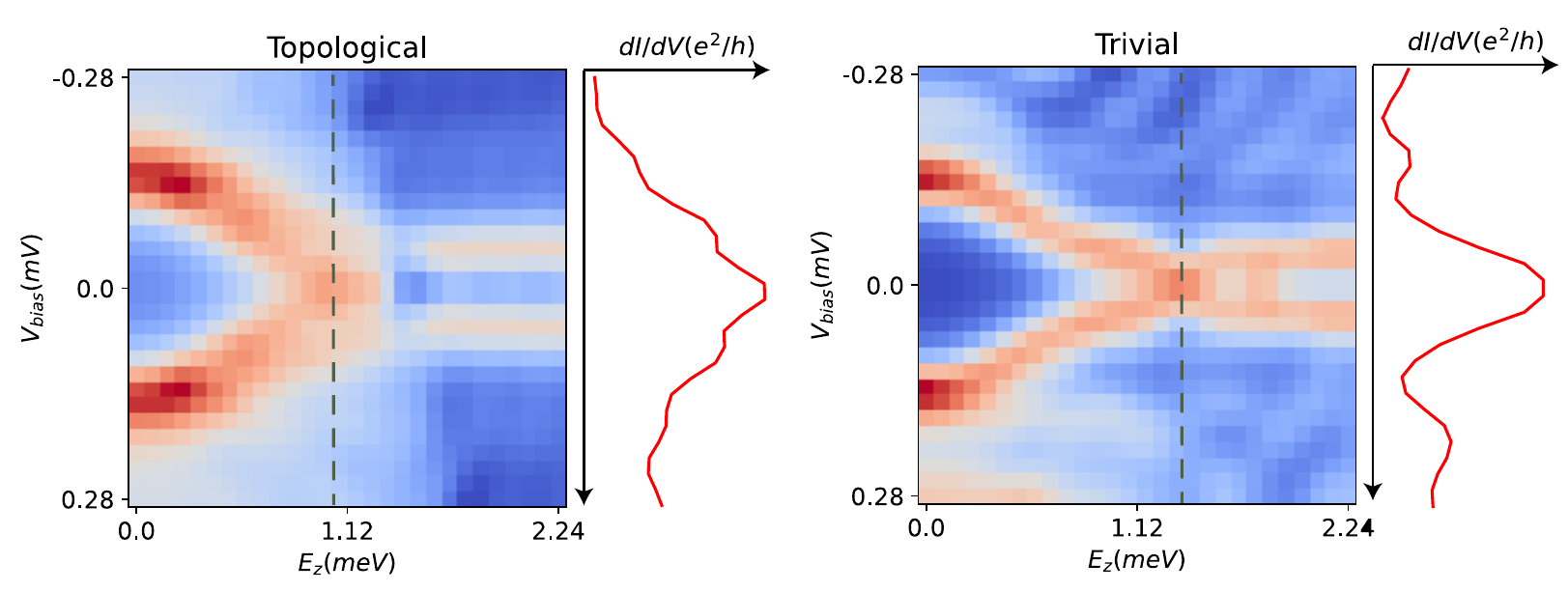}
  \caption{Example of extracting 1D curves from 2D images for topological and trivial class data.
  }
  \label{figs3.5}
\end{figure}

Fig. \ref{figs3.5} illustrates an example of the extracted 1D curves from 2D $dI/dV$ simulated signal.

\newpage
\subsection{Training with different ranges of Zeeman energy $E_z$}
Up till now our model is nicely compatible with experimental data, except the difference between $E_z=g\mu_B B/2$ in the model and the direct magnetic field $B$ in experiments. Though they are simply proportional to each other, the effective landau $g$ factor is unknown in crystals, and covers a wide range, for example from 20 to 100 \cite{jiang2022giant}. Therefore there is no straightforward way to map the experimental magnetic field into our model. Instead, for a same range of magnetic field $B\in[0,2.8T]$, we consider $g\approx 25$ and $g\approx 12.5$ separately, so the two types of corresponding $28\times 28$ pixel images have max $E_z \approx 2.24\rm{meV}$ and $1.12\rm{meV}$.

Therefore, for a same class of generated data, we have two scales of images, and we generate them separately (see Fig. \ref{figs3} below), mix them together and train jointly. In this way the machine learning process covers some message about different $g$ factors, and can perform more flexible predictions.

\begin{figure}[!htbp]
  \centering
  \includegraphics[width=0.7\textwidth]{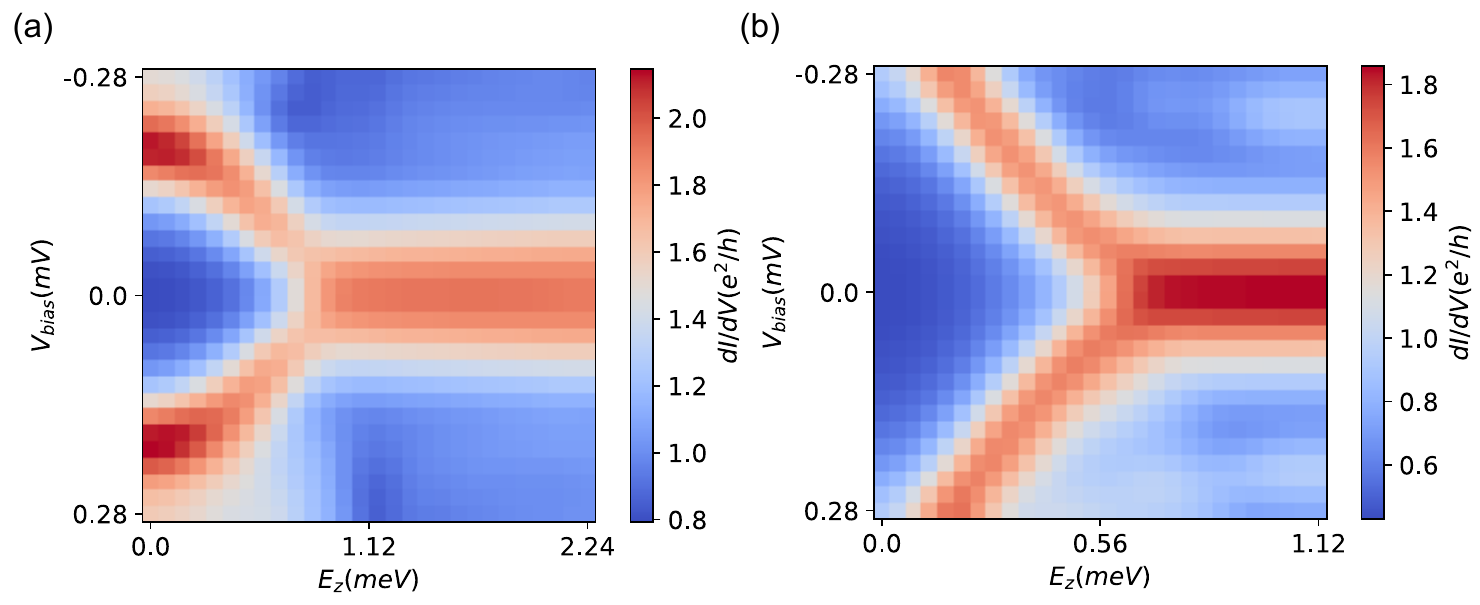}
  \caption{Example of generated data with different range of Zeeman energy $E_z$. The horizontal axis ranges are (a) $E_{z,max}=2.24\rm{meV}$ and (b) $E_{z,max}=1.12\rm{meV}$. Note that the samples are generated independently, so these are different nanowire systems.
  }
  \label{figs3}
\end{figure}

\newpage
\section{General display of dataset}
\begin{figure}[!htbp]
  \centering
  \includegraphics[width=1.0\textwidth]{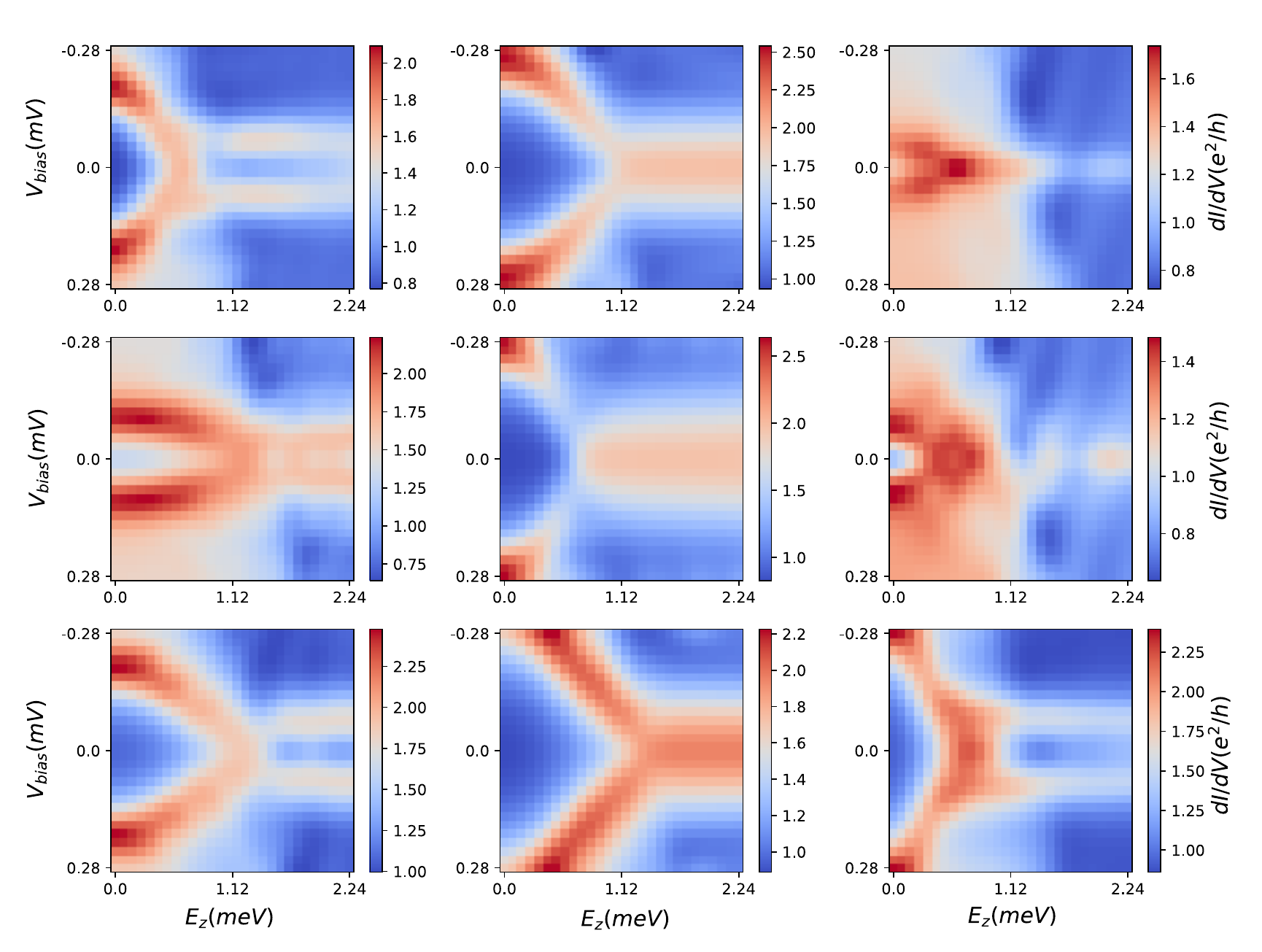}
  \caption{Example of generated 2D image data for the real Majorana Zero Mode case with minor disorder. These figures are randomly selected from the post-processed dataset.
  }
  \label{gallery1}
\end{figure}

\begin{figure}[!htbp]
  \centering
  \includegraphics[width=1.0\textwidth]{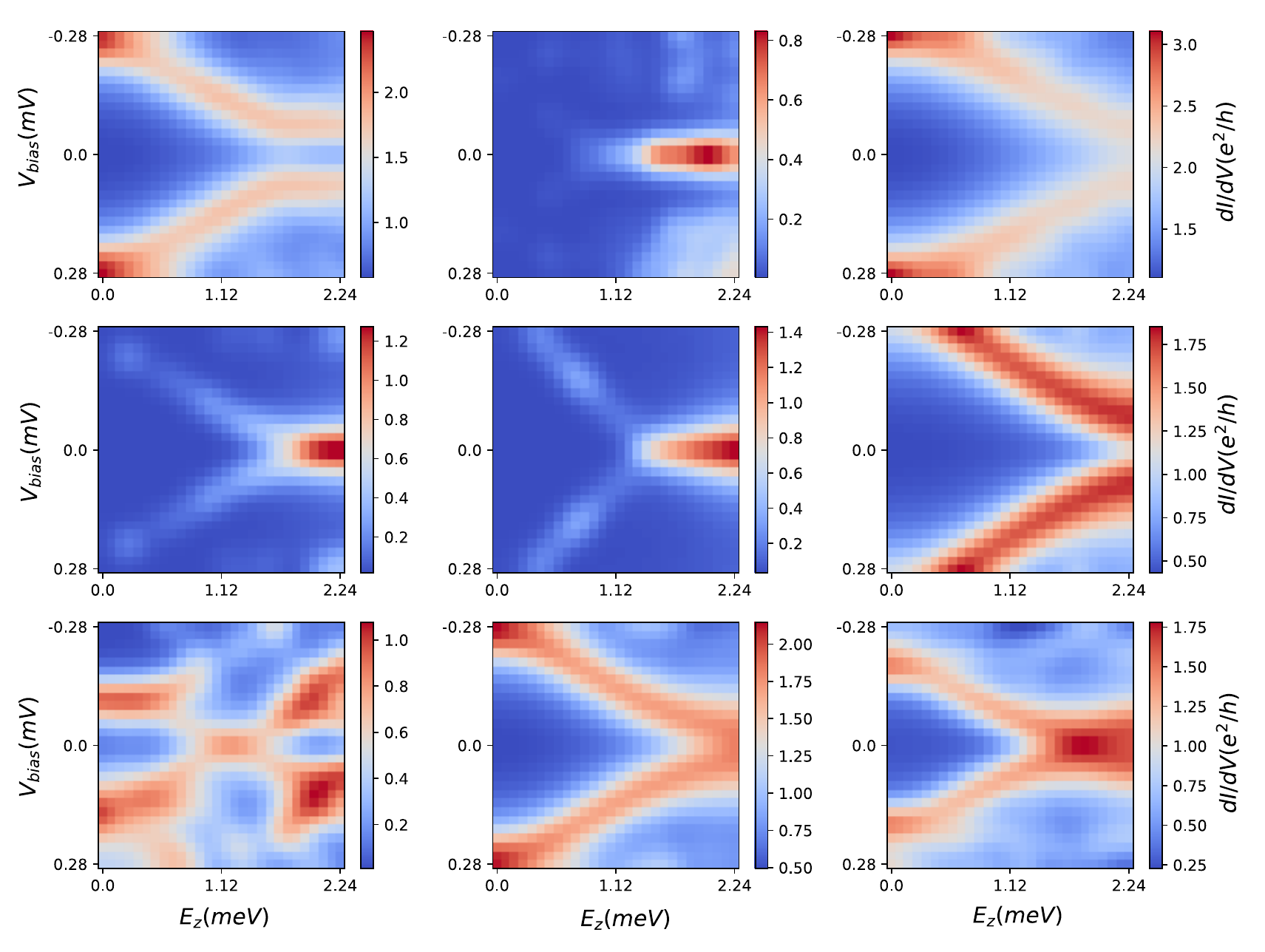}
  \caption{Example of generated 2D image data for the non-Majorana Zero Mode trivial case with incommensurate on-site potential. These figures indicate the Andreev bound state, and are randomly selected from the post-processed dataset.
  }
  \label{gallery2}
\end{figure}

\begin{figure}[!htbp]
  \centering
  \includegraphics[width=1.0\textwidth]{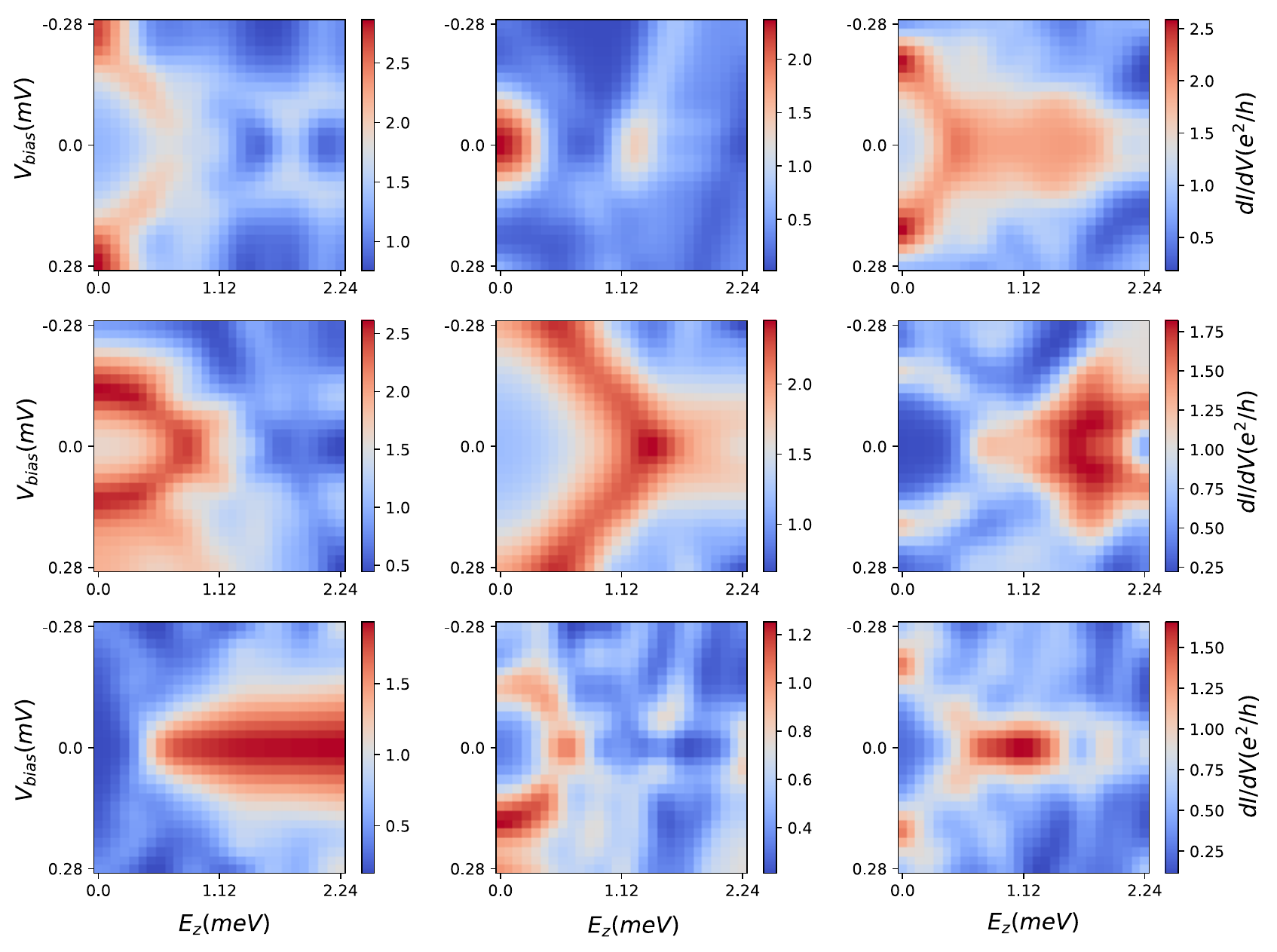}
  \caption{Example of generated 2D image data for the non-Majorana Zero Mode case with large fluctuating disorder. Topology is completely shattered in this case, and these figures are randomly selected from the post-processed dataset.
  }
  \label{gallery3}
\end{figure}

In this section we show more typical generated data for the topological MZM data and the trivial one, including the Andreev bound state data and strong disorder data. Only $E_{z,max}=2.24\rm{meV}$ is shown, because statistically $E_{z,max}=1.12\rm{meV}$ is only a zoom-in version of the former dataset. Note that for the two classes (Fig. \ref{gallery1} versus Fig. \ref{gallery2} and Fig. \ref{gallery3}), some of the figures are very difficult to directly distinguish by bare eye.

\section{Training the model}
\subsection{Anomaly detection of non-physical parameters}
Before we actually start training on the model, since the training data is generated according to our theoretical model, some extra care need to be taken to ensure we have chosen reasonable physical parameters. For example, for some choice sets of physical parameters sampled from Table \ref{table1_params}, it might not correspond to any realistic system, or simply produce unrealistic $dI/dV$ signals. 

\begin{figure}[!htbp]
  \centering
  \includegraphics[width=0.75\textwidth]{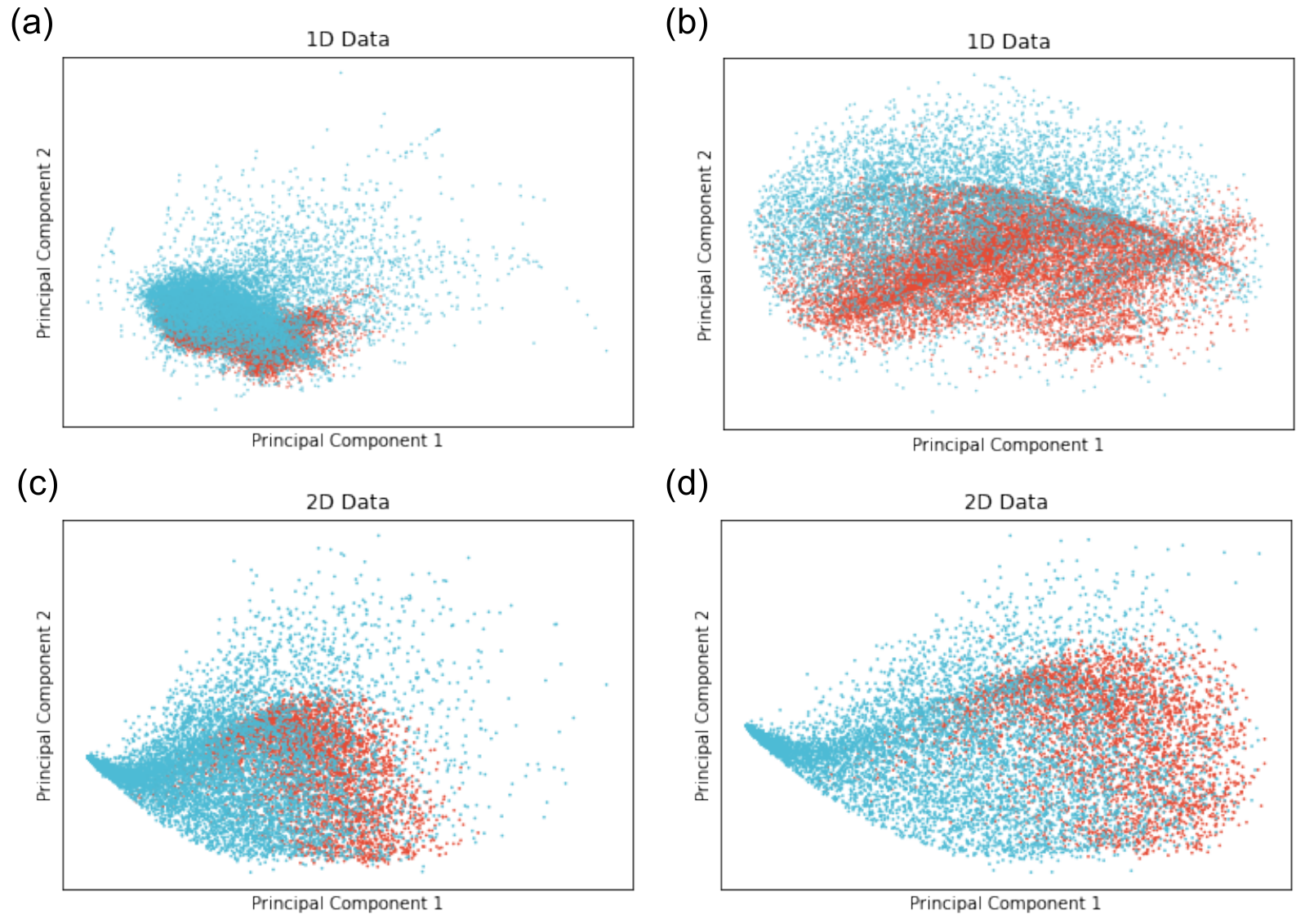}
  \caption{Anomaly detection as data processing for 1D and 2D generated data. Blue and red points correspond to trivial and topological MZM data. For better visualization, 2-principal component PCA is performed to show the pattern of data. First row shows 1D data (a) before and (b) after anomaly filter; Second row shows 2D data (c) before and (d) after anomaly filter.
  }
  \label{figs4}
\end{figure}

To avoid such unintended contamination of generated data into our model, and start training without worrying about choosing unphysical parameters, we employ a machine learning based anomaly detection algorithm to automatically filter data from our original dataset. Specifically, we use Isolation Forest method\cite{isolation_forest}, an algorithm for data anomaly detection using binary trees with a linear time complexity and a low memory requirement. 
$10\%$ of datapoints are eliminated using this algorithm, and we visualize the data distribution using Principal Component Analysis(PCA) algorithm. Figure \ref{figs4} shows scattered plots of 2-principal component PCA data distribution. After such anomaly filter most distant data points are eliminated, allowing us to focus on the main data cluster and more realistic physical parameters.

\subsection{Comparing performance of different ML methods: 2D data}
After such anomaly filter, we feed the datasets into different machine learning models and start training. The dataset is split into $75\%$ for training and $25\%$ for testing. As is mentioned in the main text, while linear classification performs poorly, more sophisticated methods like XGBoost can reach accuracy as high as $94\%$ due to hidden feature difference revealed by Topological Data Analysis.

Here we show comparison of some common machine learning techniques we applied to our training: Linear kernel Support Vector Machine (SVM), Radial basis function (RBF) kernel SVM, Random Forest (RF), XGBoost and Convolutional Neural Network (CNN).

The testing accuracy results are shown in Fig. \ref{2D classtable} below. It can be shown that non-linear classification methods can generally get $>92\%$ accuracy, while XGBoost and CNN can get the highest accuracy because of their extensive power of expressing and fitting functions. Also, testing on data without noise processing is also displayed, which gets higher accuracy due to less perturbation.
\begin{figure}[!htbp]
  \centering
  \includegraphics[width=0.55\textwidth]{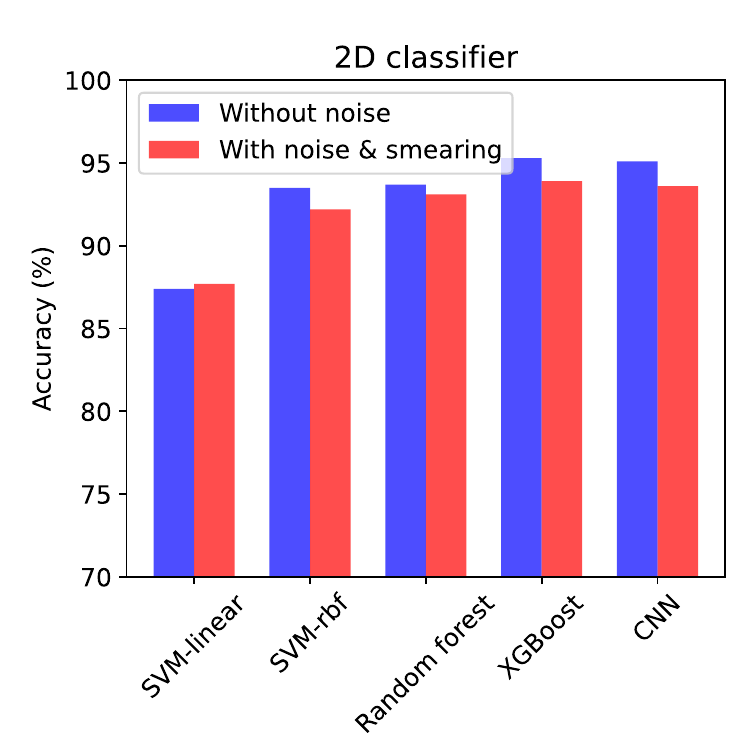}
  \caption{2D classifier testing accuracy for different models. Training results with data processed / not processed with noise and smearing are both displayed for comparison and reference.
  }
  \label{2D classtable}
\end{figure}

\subsection{Comparing performance of different ML methods: 1D data}
\begin{figure}[!htbp]
  \centering
  \includegraphics[width=0.55\textwidth]{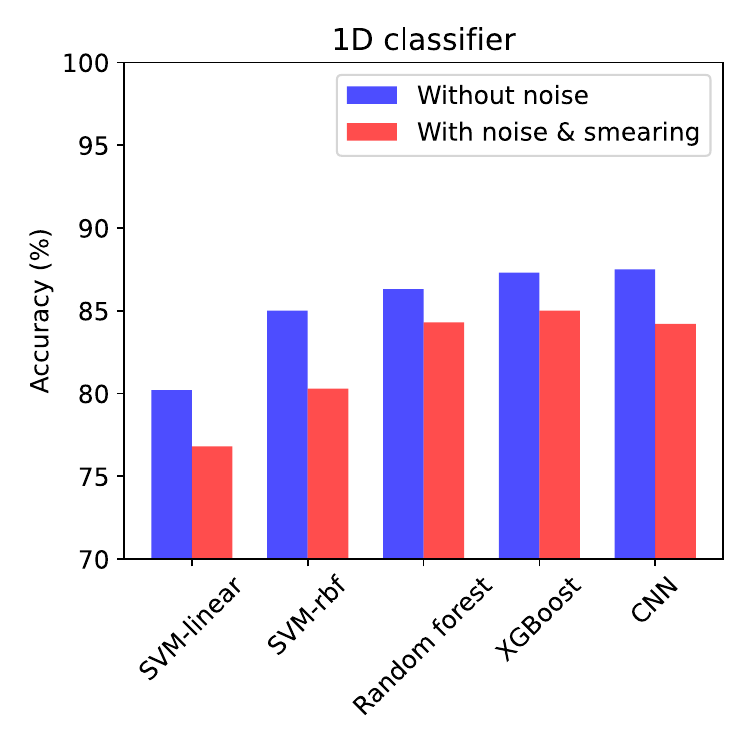}
  \caption{1D classifier testing accuracy for different models. Training results with data processed / not processed with noise and smearing are both displayed for comparison and reference.
  }
  \label{1D classtable}
\end{figure}

In parallel we also show testing results for 1D data with the same methods. As is shown in Fig. \ref{1D classtable}, 1D classifier, though losing a huge portion of information, still manages to capture $>80\%$ accuracy for non-linear classification methods.

\subsection{Model and hyperparameter settings}
In this section we document the model and hyperparameter settings for non-linear machine learning methods. For the random forest method, we choose number of estimators $n_\text{estimators}=50,200$ for 2D and 1D data correspondingly; the CNN network consists of two convolutional layers with ReLU activation function followed by 2x2 max pooling. The input layer receives a single-channel 2D image (or 1D sequence), which convolutes to 32, and then 64 output channels subsequently. The resulting feature maps are flattened and fed into a fully connected layer with 128 output features, which after undergoing a ReLU activation function, feed into the final linear layer. This latter layer generates 2 output features corresponding to a binary classification task. The model concludes its forward pass by yielding softmax probabilities for the two output classes. The learning rate is set to $r=0.001$, and we train with batch size $n_\text{batch}=64$ under 200 epoches.

For the XGBoost training, we tune the hyperparameters for 1D and 2D separately, and list them below in the table \ref{table2_hyperparams}:

\begin{table}[!htbp]
\centering
\begin{tabular}{lll} 
& 1D model & 2D model \\
\hline \hline 
Learning rate $\eta$ & 0.1 & 0.3 \\
Min split loss $\gamma$ & 5 & 1 \\
Max tree depth & 6 & 6 \\
L1 regularization $\alpha$ & 0.3 & 0 \\
L2 regularization $\lambda$ & 0 & 1 \\
\hline
\end{tabular}
\caption{Training hyperparameters for XGBoost in 1D and 2D model in our work. Other unlisted hyperparameters in the XGBoost package are not tuned, thus set to default.}
\label{table2_hyperparams}
\end{table}

\section{Testing on experimental data}
\subsection{Embedding experimental data into the classifier model}
\begin{figure}[!htbp]
  \centering
  \includegraphics[width=0.4\textwidth]{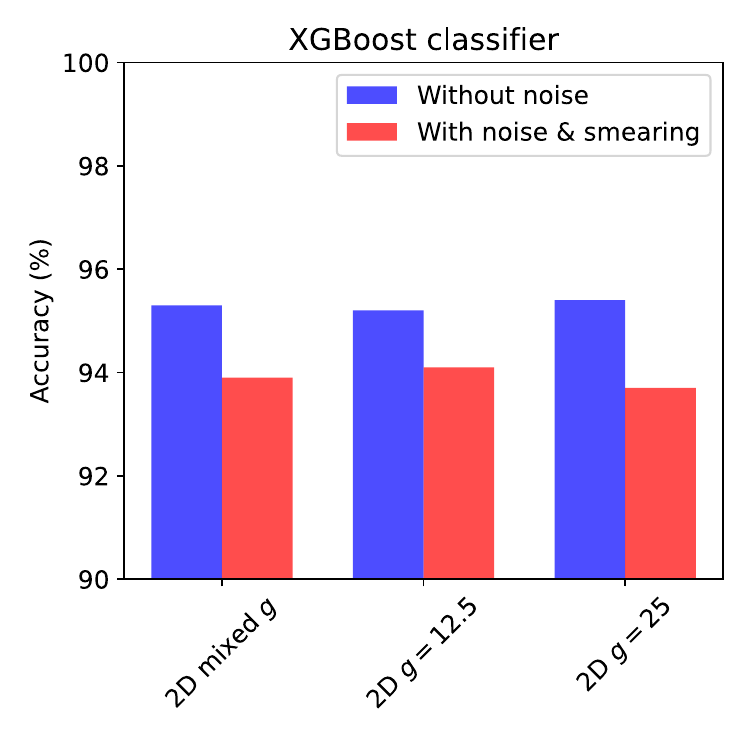}
  \caption{2D XGBoost classifier testing accuracy for datasets with different $g$ settings. Training results with data processed / not processed with noise and smearing are both displayed for comparison and reference.
  }
  \label{2D diff_g classification table}
\end{figure}

Since experimental measurement contains far denser original data than the limited pixels of our model, and the measurement range may not be compatible with our simulated ranges, we need to pre-process the experimental data into the same format as the training data before feeding it into our classifier. In our training all simulated images have $V_\text{bias}$ between $\pm0.28$ mV, and $B$ between 0 and 2.8 T with two different values of $g$. From Fig. \ref{2D diff_g classification table}, we can see that our model performs equally good on the data generated with different $g$'s. In other words, the model is robust against the scaling of $B$, i.e., the range of $B$ is not needed to be $[0, 2.8]$ T, and, instead, we can directly use what ever range the experimental data have. This means that, for any experimental image, we first crop experimental images between $V_{\text{bias}}$ within $\pm 0.28\rm{mV}$, convert the RGB pixel values into continuous conductance by linear fitting the given colorbar, and then directly discretize the image into $28\times 28$ pixels, producing a $28\times 28$ array compatible with our model dimension.

\begin{figure}[!htbp]
  \centering
  \includegraphics[width=0.85\textwidth]{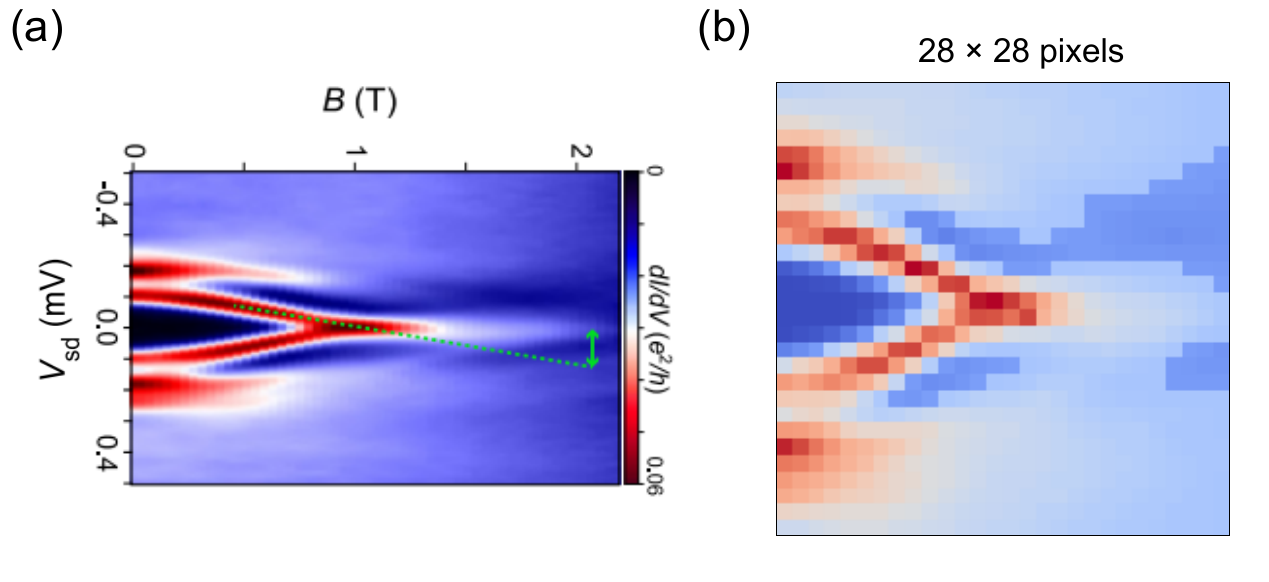}
  \caption{Embedding experimental measurements into our classifier model. (a) Original measurement taken from ref \cite{test5}; (b) Cropped results in our processing. The figure is now converted into 28 $\times$ 28 pixel image within our simulated range of parameters.
  }
  \label{embedding}
\end{figure}

It is also noticed that experimental measurements produce $dI/dV$ much lower than theoretical simulations due to strong dissipation, which is not included in our model. Due to the diversity of dissipation strength, though trivial, it is inefficient to directly apply dissipation into our simulation. Therefore at this stage we normalize experimental values to the range of $[0,4e^2/h]$, which is the double-channel conductance limit in theory. Fig. \ref{embedding} shows the typical conversion results of our framework below.

\newpage

\subsection{Testing results on 2D data}
In this subsection we show the predicted results from our model on all the 16 samples from 10 works. The first four samples from four independent works are the ones that passed our test and are discussed in the main text.

\begin{figure}[!htbp]
  \centering
  \includegraphics[width=\textwidth]{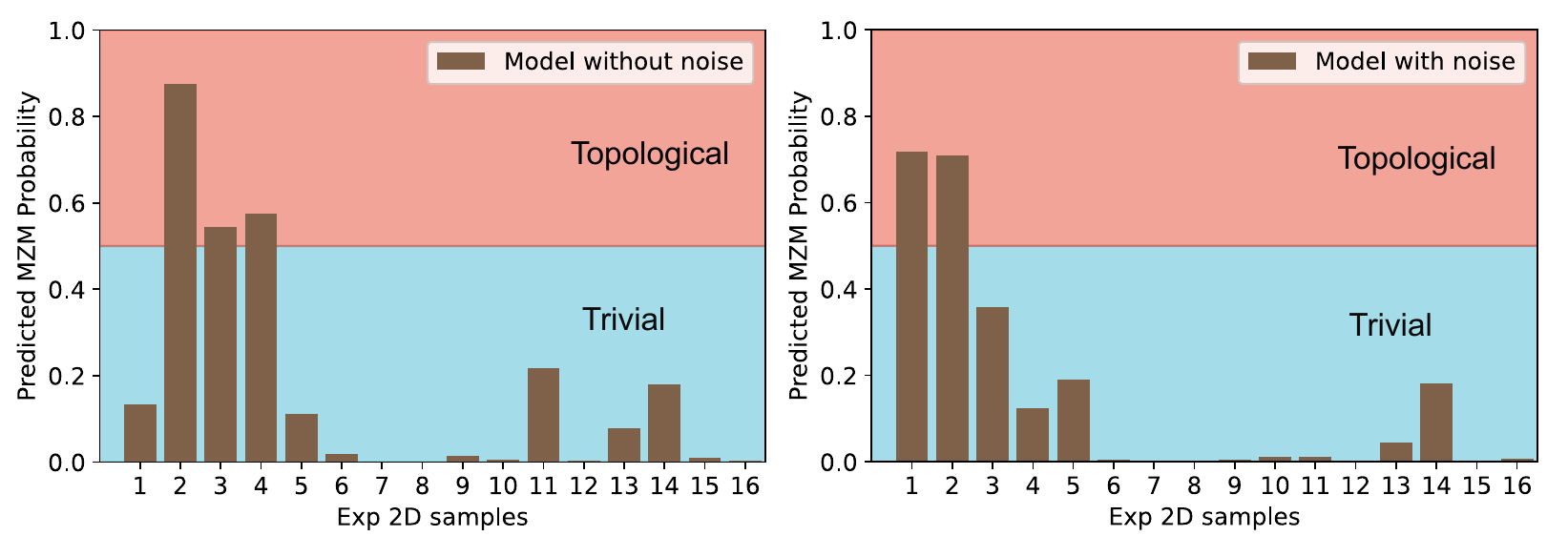}
  \caption{2D test on experimental data from ref\cite{test1,test2,test3,test4,test5,test6,test7,test8,test9,test10} under the model with/without noise.
  }
  \label{2dtest}
\end{figure}

\subsection{Testing results on 1D data}
\begin{figure}[!htbp]
  \centering
  \includegraphics[width=\textwidth]{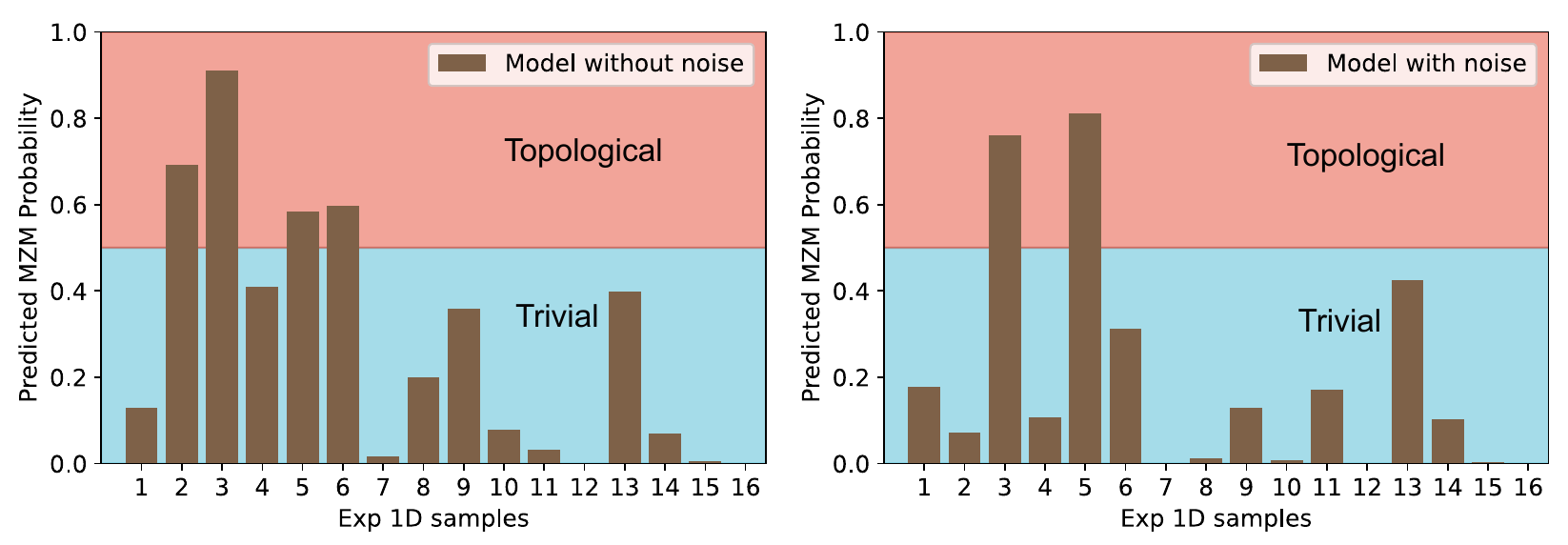}
  \caption{1D test on experimental data from ref\cite{test1,test2,test3,test4,test5,test6,test7,test8,test9,test10} under the model with/without noise.
  }
  \label{1dtest}
\end{figure}

In addition to the 2D testing result highlighted in the main text, though the 1D model is statistically less reliable, we can still embed the 1D data extracted from 2D figures into the XGBoost 1D classifier. For the same 16 measurements in 10 works\cite{test1,test2,test3,test4,test5,test6,test7,test8,test9,test10} mentioned in the main text, we can extract the 1D curves with centered zero-bias peak. Note that there might be multiple Zeeman energy $E_z$ candidates that might produce zero-bias peaks along the vertical axis in the 2D image, therefore all possible 1D curve candidates are sent into the model for evaluation. For each 2D image, i.e. each measurement, we choose the highest positive probability for containing real MZMs as the final prediction of our 1D model. The results are shown in Figure \ref{1dtest}. 

\subsection{List of experimental sample reference}
In table \ref{table3_bookmarks} below we show the bookmark for the works we considered in experimental testings. 16 figures (sample data) out of 10 works are tested in total.

\begin{table}[!htbp]
\centering
\begin{tabular}{llll} 
Sample & Work & Passed 2D test? (Y/N) \\
\hline \hline 
1 & Ref. \cite{test1} & \textbf{Y} \\
2 & Ref. \cite{test4} & \textbf{Y} \\
3 & Ref. \cite{test5} & \textbf{Y} \\
4 & Ref. \cite{test6} & \textbf{Y} \\
5 & Ref. \cite{test6} & N \\
6 & Ref. \cite{test2} & N \\
7 & Ref. \cite{test3} & N \\
8 & Ref. \cite{test3} & N \\
9 & Ref. \cite{test7} & N \\
10 & Ref. \cite{test7} & N \\
11 & Ref. \cite{test7} & N \\
12 & Ref. \cite{test8} & N \\
13 & Ref. \cite{test9} & N \\
14 & Ref. \cite{test10} & N \\
15 & Ref. \cite{test10} & N \\
16 & Ref. \cite{test10} & N \\
\hline
\end{tabular}
\caption{Bookmark of the works we considered in experimental testings.}
\label{table3_bookmarks}
\end{table}

\bibliography{bibliography.bib}. 